\documentclass[twocolumn,preprintnumbers,amsmath,amssymb,showkeys,mathtools,letterpaper]{revtex4-1}

\usepackage{graphicx}
\usepackage{upgreek}
\usepackage{amsmath}
\usepackage{amssymb}
\usepackage{dcolumn}
\usepackage{chngpage}

\usepackage{graphicx}
\usepackage{dcolumn}
\usepackage{bm}
\usepackage{amssymb} 
\usepackage{subfigure}
\usepackage{color}
\usepackage{epstopdf}
\usepackage{tabularx}
\newcolumntype{C}{>{\centering\arraybackslash}X}
\usepackage{array,booktabs}

\begin{document}

\preprint{{\it J. Chem. Phys.}}

\title{NMR Spin-Rotation Relaxation and Diffusion of Methane}

\author{P. M. Singer}\email{ps41@rice.edu}
\author{D. Asthagiri}\email{dna6@rice.edu}
\author{W. G. Chapman}
\author{G. J. Hirasaki}
\affiliation{Department of Chemical and Biomolecular Engineering, Rice University, 6100 Main St., Houston, TX 77005, USA}

\date{\today}


\begin{abstract}
	
The translational-diffusion coefficient $D_T$ and the spin-rotation contribution to the $^1$H NMR relaxation time $T_{1J}$ for methane (CH$_4$) are investigated using MD (molecular dynamics) simulations, over a wide range of densities $\rho$ and temperatures $T$, spanning the liquid, supercritical, and gas phases. The simulated $D_T$ agree well with measurements, without any adjustable parameters in the interpretation of the simulations. A minimization technique is developed to compute the angular-velocity for non-rigid spherical molecules, which is used to simulate the autocorrelation function $G_{\!J}(t)$ for spin-rotation interactions. With increasing $D_T$ (i.e. decreasing $\rho$), $G_{\!J}(t)$ shows increasing deviations from the single-exponential decay predicted by the Langevin theory for hard spheres, and the deviations are quantified using inverse Laplace transforms of $G_{\!J}(t)$. $T_{1J}$ is derived from $G_{\!J}(t)$ using the kinetic model ``{\it km}" for gases ($T_{1J}^{km}$), and the diffusion model ``{\it dm}" for liquids ($T_{1J}^{dm}$). $T_{1J}^{km}$ shows better agreement with $T_1$ measurements at higher $D_T$, while $T_{1J}^{dm}$ shows better agreement with $T_1$ measurements at lower $D_T$. $T_{1J}^{km}$ is shown to dominate over the MD simulated $^1$H-$^1$H dipole-dipole relaxation $T_{1RT}$ at high $D_T$, while the opposite is found at low $D_T$. At high $D_T$, the simulated spin-rotation correlation-time $\tau_J$ agrees with the kinetic collision time $\tau_K$ for gases, from which a new relation $1/T_{1J}^{km} \propto D_T$ is inferred, without any adjustable parameters.

\end{abstract}

\maketitle

\section{Introduction}\label{sc:Introduction}

Theoretical and experimental investigations into $^1$H NMR (nuclear magnetic resonance) spin-rotation relaxation $T_{1J}$ and translational-diffusion $D_T$ of methane (CH$_4$), deutero-derivatives (CH$_{4-n}$D$_n$), and halide-derivatives (CH$_{4-n-m}$F$_n$Cl$_m$) dates back over 50 years \cite{gutowsky:prl1961,hubbard:pr1963,bloom:cjp1967a,bloom:cjp1967b,mcclung:JCP1969,dong:cjp1970,dawson:aiche1970,mcclung:JCP1971,gerritsma:ph1971a,gerritsma:ph1971b,oosting:ph1971a,oosting:ph1971b,rajan:jmr1974,mcclung:AMRIP1977,mcclung:emr2007}. Likewise, the theoretical and experimental investigations into molecular-beam magnetic-resonance of methane and its derivatives dates back over 50 years \cite{cederberg:pr1964,anderson:pr1966,yi:pr1968,ozier:pr1968,yi:jcp1971}, from which came (among many other things) the coupling constants used to interpret $T_{1J}$.

More recently, the influence of dissolved methane on the $T_{1}$ and $D_T$ of complex crude-oils and hydrocarbon mixtures has been investigated \cite{zhang:spwla2002,lo:SPE2002,hurlimann:petro2009,yang:petro2012}, which reveal the important influence of the spin-rotation contribution from methane. Of particular significance is the well established spin-rotation component for methane in the fast-motion regime \cite{rajan:jmr1974,lo:SPE2002}:
\begin{align}
\frac{1}{T_{1J}^A} = A \frac{T^{3/2}}{\rho},
\label{eq:T1lo}
\end{align}
where $A$ is an empirically derived constant. Eq. \ref{eq:T1lo} states that the relaxation rate $1/T_{1J}$ increases with temperature $T$, i.e. $T_{1J}$ decreases with $T$. This is in stark contrast to the $^1$H-$^1$H dipole-dipole relaxation $T_{1RT}$ which increases with $T$, for all hydrocarbons, including methane. Also of interest recently is the influence of pore confinement on the NMR response of methane \cite{straley:spwla1997,sigal:petro2011,kausik:spe2011,tinni:spe2014,wang:ef2014,papaioannou:PRA2015,sigal:spej2015,kausik:petro2016,valori:ef2017,tinni:petro2018}, which has practical applications for characterizing the light hydrocarbons in the organic nano-pores of kerogen and bitumen in organic-rich shale. One of the current mysteries is why the $T_{1S}/T_{2S}$ ratio for surface-relaxation of methane in organic-shale is typically $T_{1S}/T_{2S} \lesssim 2$, while for higher-order alkanes it is typically higher $T_{1S}/T_{2S} \gtrsim 4$. This has practical applications for separating the NMR response of light hydrocarbon from water in organic-shale, and for determining the hydrocarbon saturation in the organic-shale reservoir.

In order to properly characterize NMR relaxation of methane in the bulk and under nano-pore confinement, one must first separate the $^1$H spin-rotation relaxation $T_{1J}$ from $^1$H-$^1$H dipole-dipole relaxation $T_{1RT}$. Traditionally this has been done by partially deuterating CH$_4$ to, for instance CHD$_3$, which dramatically reduces the dipole-dipole contribution. However, this has the drawback of turning a spherical molecule (CH$_4$) with one principle moment of inertia, into a symmetric-top molecule (CHD$_3$) with two principle moments of inertia. The theory of spin-rotation relaxation for symmetric-top molecules is much more complex than for spherical molecules, thereby making comparisons with measurements more complex.

MD (molecular dynamics) simulations provide an ideal tool for separating $^1$H NMR relaxation mechanisms. As already shown for liquid-state $n$-alkanes in Ref. \cite{singer:jmr2017}, MD simulations can naturally separate in{\it tra}molecular $T_{1R}$ from in{\it ter}molecular $T_{1T}$ $^1$H-$^1$H dipole-dipole relaxation, without deuteration, and without any adjustable parameters in the interpretation of the simulations. Such simulations yield unique insights into (a) the relative strengths of in{\it tra}molecular versus in{\it ter}molecular relaxation, (b) the influence of internal motions on the molecular dynamics of non-rigid molecules, and (c) the validity of traditional hard-sphere models \cite{bloembergen:pr1948,torrey:pr1953} for different molecular geometries. 

In this report we simulate spin-rotation relaxation. In Section \ref{ssc:MolecularSimulation}, we develop a technique to determine the autocorrelation function for angular-velocity $G_{\!J}(t)$ of non-rigid spherical molecules. In Section \ref{ssc:SpinRotationRelaxation} we interpret $G_{\!J}(t)$ to yield the spin-rotation relaxation using the kinetic model and the diffusion model. In Section \ref{ssc:SimulationVersusMeasurement} we compare simulation versus measurement for spin-rotation relaxation and translational diffusion. In Section \ref{ssc:CorrelationTimes} we compare and analyze the correlation times for the different relaxation mechanisms. In Section \ref{ssc:KineticModel} we propose a new kinetic model to account for Eq. \ref{eq:T1lo}.


\section{Methodology} \label{sc:Methodology}

\subsection{Molecular simulation} \label{ssc:MolecularSimulation}

The MD simulations were performed using NAMD \cite{namd} version 2.11. Methane was modeled using the CHARMM General Force field (CGenFF) \cite{cgenff}. The protocol for setting-up the initial simulation configuration was exactly as before \cite{singer:jmr2017}. As before, we created the initial simulation system by packing $N$ copies of the molecule into a cube of volume $L^3$ using the Packmol program \cite{packmol}. The volume was chosen such that the number density $N/V$ corresponds to the experimentally determined number density at the specified temperatures listed in Table \ref{tb:1}. The simulation approach for these systems using NAMD was as before \cite{singer:jmr2017}. 

The angular-velocity computation went as follows. From simulations, for each atom in a molecule, we have a site velocity $\bm{v}_i$. Without loss of generality, we assume $\bm{v}_i$ is relative to the velocity of the center of mass. By the definition of angular velocity, we have 
\begin{equation}
	\bm{\omega} \times \bm{r}_i = \bm{v}_i.
\end{equation}
But a direct application of the above equation cannot be used to calculate $\bm{\omega}$ because the matrix equation is singular. We therefore define 
\begin{equation}
\mathcal{L} = \sum_i | \bm{v}_i - \bm{\omega} \times \bm{r}_i|^2.
\end{equation}
We minimize $\mathcal{L}$ with respect to the $x, y, z$-components of $\bm{\omega}$, giving the following matrix equation:
\begin{multline}
\begin{pmatrix}
	\sum_{i} z_i^2 + y_i^2  & -\sum_i x_i y_i           & -\sum x_i z_i \\
	-\sum_i x_i y_i            & \sum_i x_i^2 + z_i^2  & -\sum y_i z_i \\
	-\sum_i x_i z_i            & -\sum y_i z_i               & \sum_i x_i^2 + y_i^2
\end{pmatrix}
\begin{pmatrix}
	\omega_x \\
	\omega_y \\
	\omega_z 
\end{pmatrix}
= \\
\begin{pmatrix}
	\sum_i (\bm{r}_i \times \bm{v}_i)_x \\
	\sum_i (\bm{r}_i \times \bm{v}_i)_y \\
	\sum_i (\bm{r}_i \times \bm{v}_i)_z 
\end{pmatrix}
\end{multline}
In tensor notation, we can write this more compactly as
\begin{equation}
r_j \omega_p r_q \epsilon^{ljk} \epsilon_{pqk} = r_j v_k \epsilon_{jkl},
\end{equation}
summed over all sites. 

We constructed the above matrix for each molecule and solved for $\omega_x, \omega_y, \omega_z$. This calculation is repeated for all the frames in the trajectory. Then we calculated the autocorrelation of each component. Since the system is isotropic, all the component relaxations are the same. The above calculation is repeated for other molecules and the results averaged.

\renewcommand{\arraystretch}{1.2}
\begin{table}[!ht]
	\begin{tabular}{cccccccc}
		\hline
		
		Phase& $T$ & $P$ & $\rho$ & $D_T$ & $T_1$&  $\eta$ & $L$ \\ 
		& (K) & (bar) & (g/cm$^3$)  & (10$^{-9}$m$^2\!$/s)  & (s)&  (cP) & (${\rm \AA}$)\\ 
		pred. & meas. & pred. & meas.  & meas. & meas. &  pred. & \\ 
		\hline
		L	&	90.9	&	0.120	&	0.451	&	2.52	&	9.7	&	0.192	&	31.15	\\
		L	&	105.3	&	0.578	&	0.432	&	4.35	&	14.5	&	0.134	&	31.62	\\
		L	&	125.0	&	2.69	&	0.402	&	7.76	&	18.8	&	0.089	&	32.37	\\
		L	&	142.9	&	7.41	&	0.372	&	11.8	&	20.2	&	0.064	&	33.23	\\
		S	&	194.8	&	389.8	&	0.359	&	18.0	&	16.4	&	0.056	&	33.63	\\
		S	&	194.8	&	141.0	&	0.303	&	25.5	&	14.9	&	0.037	&	35.56	\\
		S	&	194.8	&	71.24	&	0.255	&	34.0	&	12.8	&	0.027	&	37.69	\\
		S	&	194.8	&	59.76	&	0.230	&	39.4	&	11.6	&	0.024	&	38.96	\\
		S	&	194.8	&	52.53	&	0.173	&	56.2	&	9.05	&	0.017	&	43.35	\\
		S	&	298.2	&	349.1	&	0.232	&	56.5	&	7.11	&	0.027	&	38.89	\\
		S	&	298.2	&	250.0	&	0.188	&	74.5	&	5.95	&	0.022	&	41.69	\\
		S	&	194.8	&	46.53	&	0.089	&	116.1	&	4.77	&	0.010	&	40.20	\\
		S	&	298.2	&	151.1	&	0.120	&	120.4	&	3.61	&	0.016	&	48.50	\\
		S	&	273.2	&	99.46	&	0.089	&	160	&	3.04	&	0.014	&	40.14	\\
		S	&	273.2	&	82.56	&	0.072	&	200	&	2.44	&	0.013	&	43.07	\\
		S	&	273.2	&	64.68	&	0.054	&	264	&	1.74	&	0.012	&	47.41	\\
		G	&	273.2	&	45.29	&	0.036	&	404	&	1.21	&	0.011	&	54.51	\\
		S	&	307.7	&	51.73	&	0.035	&	439	&	0.96	&	0.012	&	54.76	\\

		\hline
	\end{tabular}
	\caption{List of MD simulated state points, chosen to coincide with measured (meas.) methane data taken from Ref. \cite{gerritsma:ph1971a,gerritsma:ph1971b,oosting:ph1971a,oosting:ph1971b}, including phase (Liquid, Supercritical, or, Gas), temperature ($T$), pressure ($P$), density ($\rho$), translational diffusion ($D_T$), $^1$H NMR relaxation time ($T_1$), viscosity ($\eta$), and MD cube size $L$, in order of increasing $D_T$. Predicted quantities (pred.) are taken from NIST database. Critical points for methane are $T_{cr} = 190.6$ K and $P_{cr} = 46$ bar. \label{tb:1} }
\end{table}

The translational-diffusion coefficient $D_T$ was derived in a similar fashion to Ref. \cite{singer:jmr2017}. The linear slope of the mean-squared displacement was computed in the interval between 5 ps $\leftrightarrow$ 10 ps. The periodic-boundary correction term \cite{yeh:jpcb2004,kremer:jcp93} was derived using viscosity $\eta$ and box size $L$, both listed in Table \ref{tb:1}. The correction term resulted in a 5 $\leftrightarrow$ 10 \% boost in diffusion coefficient.

\subsection{Spin-rotation relaxation} \label{ssc:SpinRotationRelaxation}

The Hamiltonian for the spin-rotation interaction is given by the following \cite{cederberg:pr1964,anderson:pr1966,yi:pr1968,ozier:pr1968,yi:jcp1971}:
\begin{equation}
\mathcal{H}_{J} = \hbar \sum_{i=1}^4 {\bf I}_i\cdot{\bf C}_i\cdot{\bf J}.
\label{eq:HamiltonianJ}
\end{equation}
${\bf I}_i$ are the four $^1$H nuclear spins on the methane molecule. ${\bf J}$ is the angular momentum of the molecule, which is related to the angular velocity 
${\bm \omega}$ by the following $\hbar {\bf J} = I {\bm \omega}$, where $I = 5.33 \times 10^{-47}$ kg m$^2$ \cite{herzberg:book} is the moment of inertia for methane. The coupling tensor ${\bf C}$ has two principle components $C_{\perp}$ and $C_{\scriptscriptstyle{\parallel}}$, which are proportional to the magnetic field (per unit $J$) generated at a $^1$H by rotations about an axis $\perp$ ($\parallel$) to the C-H bond axis, respectively. The most practical formulation separates $\mathcal{H}_{J}$ into a scalar component proportional to the average coupling constant $ C_a = \left(2C_{\perp} + C_{\scriptscriptstyle{\parallel}}\right)\!/3$, and a tensor component proportional to the diagonal coupling constant $ C_d = \left(C_{\perp} - C_{\scriptscriptstyle{\parallel}}\right)$. The accepted values of the coupling constants, and the ones used here, are \cite{yi:jcp1971}: $C_a/2\pi = 10.4 \pm 0.1 \,{\rm kHz}$, $ C_d/2\pi = 18.5 \pm 0.5 \,{\rm kHz}$ (which are the experimentally determined quantities), from which the following can be inferred $C_{\perp}/2\pi = 16.6 \pm 0.3\,{\rm kHz}$, $C_{\scriptscriptstyle{\parallel}}/2\pi = -1.9 \pm 0.2\,{\rm kHz}$.

The expression for the autocorrelation function $G_{\!J}^{lk}\!(t) $ used to determine $T_{1J}$ for liquids \cite{hubbard:pr1963} and gases \cite{bloom:cjp1967a,bloom:cjp1967b,dong:cjp1970} can be summarized as follows (in units of s$^{-2}$):
\begin{multline}
G_{\!J}^{lk}\!(t) = \frac{I^2}{2 \hbar^2}  \bigl(C_a^2 \left< \omega_l(t\!+\!\tau) \, \omega_{k}(\tau)  \right>_{\! \tau} \, + \\  \tfrac{2}{9}C_d^2 \!\!\!\! \sum\limits_{m,n,o,p} \!\!\!\!\alpha_{l,m,n}^{k,o,p} \left< Y_{2}^{n}\!(t\!+\!\tau) \,\omega_{m}\!(t\!+\!\tau) \, Y_{2}^{p}\!(\tau) \,\omega_{o}\!(\tau)  \right>_{\!\tau} \bigr). \label{eq:Gt}
\end{multline}
The first term in Eq. \ref{eq:Gt} is the scalar term, while the second term is the tensor term. The more general expression also includes a cross-term proportional to $C_a C_d$, however it reduces to zero for spherically-symmetric molecules such as methane. The terms $\omega_k(t)$ is the angular velocity around the $k$ axis at time $t$, and $Y_2^m(t)$ is the spherical harmonic of rank 2 and order $m$ at time $t$. $\alpha_{l,m,n}^{k,o,p}$ is composed of two sets of 3-$j$ symbols defined in \cite{hubbard:pr1963}.

\begin{figure}[!ht]
	\begin{center}
		\includegraphics[width=0.9\columnwidth,trim={0.2cm 0cm 0.4cm 0.4cm},clip]{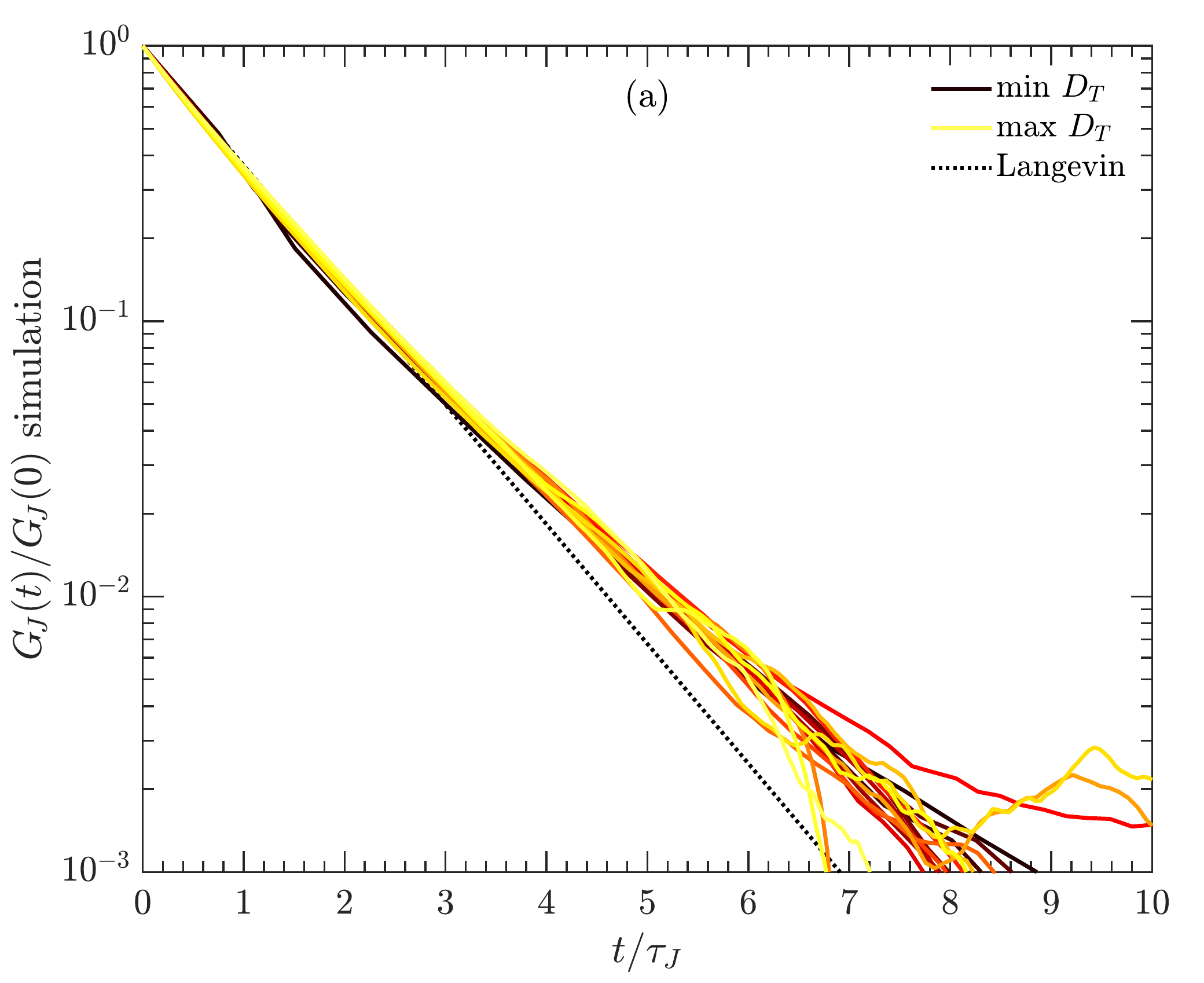}
		\includegraphics[width=0.9\columnwidth,trim={0.2cm 0cm 0.4cm 0.4cm},clip]{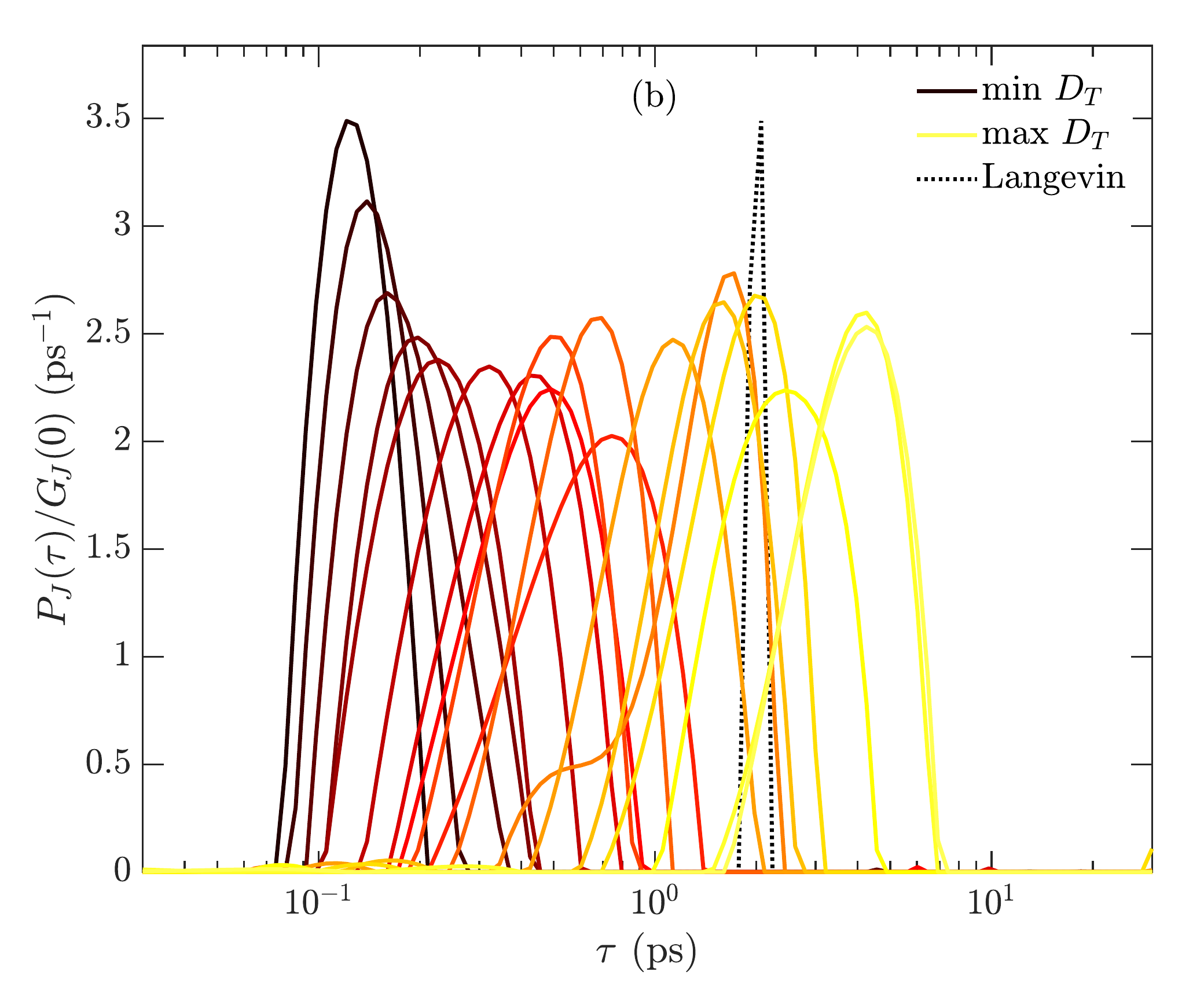}
	\end{center}
	\caption{(a) MD simulations of the autocorrelation function $G_{\!J}(t)$ for spin-rotation  interactions using Eq. \ref{eq:Gt1}, colored in order of increasing $D_T$. The $y$-axis has been normalized by zero time value $G_{\!J}(0)$ in Eq. \ref{eq:Gt0}, and the $x$-axis has been normalized by correlation time $\tau_{J}$ in Eq. \ref{eq:TauJ}. The dotted line is Langevin prediction in Eq. \ref{eq:Langevin}. (b) Probability distribution function $P_J(\tau)$ of spin-rotational correlation time $\tau$ derived from the inverse Laplace transform (Eq. \ref{eq:ILT}) of the $G_{\!J}(t)$ simulations in (a). Also shown is the Langevin prediction $G_{\!J}(t)$ from Eq. \ref{eq:Langevin}, generated with an arbitrarily chosen value of $\tau_J = $ 2 ps. The $y$-axis has been divided by $G_{\!J}(0)$, which normalizes the area of the distributions to unity (except for the Langevin model).} 
	\label{fg:GtDecay}
\end{figure}

The tensor term greatly simplifies provided the rotational motion $Y_2^m(t)$ is independent of the angular velocity $\omega_k(t)$. In such cases the bracketed term separates as $ \left< Y_{2}^{n}(t\!+\!\tau)  \, Y_{2}^{p}(\tau)\right>_{\! \tau}\left<\omega_{m}\!(t\!+\!\tau) \,\omega_{o}\!(\tau) \right>_{\! \tau}$. Traditional theories state the following expressions for hard spheres:
\begin{align}
\left< \omega_l(t\!+\!\tau)\, \omega_{k}(\tau) \right>_{\! \tau} &=  \delta_{l,k}\frac{ k T}{I} \exp\bigl(-t/\tau_J\bigr),\label{eq:Langevin} \\
\left< Y_{2}^{l*}(t\!+\!\tau) \, Y_{2}^{k}(\tau) \right>_{\! \tau} &= \delta_{l,k} \exp\bigl(-t/\tau_R\bigr). \label{eq:BPP}
\end{align}
The angular-velocity autocorrelation function in Eq. \ref{eq:Langevin} is derived from the Langevin model with ``friction time" $\tau_J$ \cite{mcconnell:book}. The orientation autocorrelation function in Eq. \ref{eq:BPP} is the Deybe model with rotational correlation-time $\tau_R$ \cite{bloembergen:pr1948}. The orientation and angular-velocity autocorrelations can be separated provided either $\tau_J \gg \tau_R$ (gases) or $\tau_J \ll \tau_R$ (liquids) \cite{hubbard:pr1963}. In the case of liquids ($\tau_J \ll \tau_R$), this has the effect of making $\left< Y_{2}^{l*}(t\!+\!\tau) \, Y_{2}^{k}(\tau) \right>_{\!\! \tau} = \delta_{l,k}$ in Eq. \ref{eq:Gt}. Based on comparison with measurements, the relation $\left< Y_{2}^{l*}(t\!+\!\tau) \, Y_{2}^{k}(\tau) \right>_{\!\! \tau} = \delta_{l,k}$ is also used in the case of gases \cite{dong:cjp1970,rajan:jmr1974,mcclung:emr2007}. In such cases, Eq. \ref{eq:Gt} simplifies to the following expression which is used in the MD simulations:
\begin{equation}
G_{\!J}(t) = \frac{I^2}{2 \hbar^2}  \left(C_a^2 + \alpha \tfrac{2}{9}C_d^2\right)  \left< \omega_k(t\!+\!\tau) \, \omega_{k}(\tau)  \right>_{\! \tau}  \label{eq:Gt1}
\end{equation}
The resulting expression $G_{\!J}(t)$ is independent of the direction $k$ in $G_{\!J}^{kk}\!(t)$, therefore the superscript is removed for clarity. The MD simulations compute all three $k$ directions independently, and the average is then taken to improve the signal to noise ratio. The simulation results for $G_{\!J}(t)$ are shown in Fig. \ref{fg:GtDecay}(a), where both $x$ and $y$ axes have been normalized for better comparison of the functional form of the decay between the different states. The normalization also allows for comparison with the Langevin model in Eq. \ref{eq:Langevin}.

A significant parameter in the analysis is the autocorrelation
at $t = 0$, which is given by the following expression:
\begin{multline}
G_{\!J}(0) = \frac{I^2}{2 \hbar^2}  \left(C_a^2 +  \alpha\tfrac{2}{9}C_d^2\right)  \left<  \omega^2_{k}(\tau)  \right>_{\! \tau}= \\    \frac{I kT}{2 \hbar^2}  \left(C_a^2 +  \alpha\tfrac{2}{9}C_d^2\right) = \frac{1}{2}\Delta\omega_{\alpha}^2,
\label{eq:Gt0}
\end{multline}
where the first equality is directly from Eq. \ref{eq:Gt1}. The second equality uses the time zero expression $\left<  \omega^2_{k}(\tau) \right>_{\! \tau}  = kT/I $ from Eq. \ref{eq:Langevin}, which was verified from MD simulations to be within $\pm$1.5 \% over the entire temperature range of interest. The third equality defines the second moment $\Delta\omega_{\alpha}^2$ (i.e. strength) of the spin-rotation interaction \cite{cowan:book}, given by:
\begin{align}
\Delta\omega_{km}^2 &= \frac{I kT}{\hbar^2} \! \left(C_a^2 +  \tfrac{4}{45}C_d^2\right) \,\,[{\rm k.m.}, \alpha = \tfrac{2}{5}], \label{eq:SMkm} \\
\Delta\omega_{dm}^2 &= \frac{I kT}{\hbar^2} \! \left(C_a^2 +  \tfrac{2}{9}C_d^2\right) \,\,[{\rm d.m.}, \alpha = 1]. \label{eq:SMdm}
\end{align}
The only free parameter is $\alpha$, which is given by $\alpha = 1$ in the diffusion model ``{\it dm}" for liquids \cite{hubbard:pr1963}, or by $\alpha = 2/5$ in the kinetic model ``{\it km}" for gases \cite{dong:cjp1970,rajan:jmr1974}. In describing the kinetic model, we have adopted the exact relation for the scalar term $\hbar^2\!\left<J(J+1)\right>\!/3= I kT$, and for the tensor term we assume (without loss in accuracy) that $\left<(2J-1)(2J+3)\right>/4\left<J(J+1)\right> \simeq 1$ in the classical limit $J \gg 1$ \cite{bloom:cjp1967b}. The reason for $\alpha = 2/5$ in the kinetic model for gases is that the oscillatory terms $\Delta J \neq 0$ do not contribute (giving rise to $\times 1/5$), and from the statistical independence of $\omega_k(t)$ and $Y_2^m(t)$ (giving rise to $\times 2$) \cite{bloom:cjp1967b}. 

The next quantity of interest is the spin-rotation correlation time $\tau_J$ determined from the MD simulations, which is determined from the integral of the normalized $G_{\!J}(t)$ as such \cite{cowan:book}:
\begin{equation}
\tau_{J} = \frac{1}{G_{\!J}(0)}\int_{0}^{\infty}\!G_{\!J}(t)\, dt.
\label{eq:TauJ}
\end{equation}
The NMR relaxation times are then derived from the spectral density $J_J^{\alpha}(\omega)$ \cite{hubbard:pr1963}:
\begin{align}
J_J^{\alpha}(\omega) &= 2\int_{0}^{\infty}\!G_{\!J}(t)\cos\left(\omega t\right) dt, \nonumber \\
\frac{1}{T_{1J}^{\alpha}}&= 2 J_J^{\alpha}\!\left(\omega_0\right), \label{eq:T1spectral}\\
\frac{1}{T_{2J}^{\alpha}}&= J_J^{\alpha}\!\left(0\right) + J_J^{\alpha}\!\left(\omega_0\right),
\nonumber
\end{align}
where $\omega_0 = \gamma B_0$ is the Larmor frequency for $^1$H. Given the short correlation times $\tau_J \sim 1$ ps, and given typical Larmor frequencies $\omega_0/2\pi < 500 $ MHz, it is clearly the case that $\omega_0 \tau_{J} \ll 1$, i.e. the fast-motion regime applies. In such cases $ J_J^{\alpha}(0) = J_J^{\alpha}(\omega_0)  = \Delta\omega_{\alpha}^2 \tau_{J}$, and therefore $T_{1J}^{\alpha} = T_{2J}^{\alpha}$. Using the relation $1/T_{1J}^{\alpha} = 2J_J^{\alpha}(0) = 2 \Delta\omega_{\alpha}^2  \tau_{J}$ results in the final expressions:
\begin{align}
\frac{1}{T_{1J}^{km}}&= \frac{2I kT}{\hbar^2} \! \left(C_a^2 +  \tfrac{4}{45}C_d^2\right) \tau_{J}, \label{eq:T1Jkm} \\
\frac{1}{T_{1J}^{dm}} &= \frac{2I kT}{\hbar^2} \! \left(C_a^2 +  \tfrac{2}{9}C_d^2\right) \tau_{J}  \label{eq:T1Jdm}.
\end{align}
Note that $T_{1J}^{dm}$ is often expressed in terms of $C_{\perp}$ and $C_{\scriptscriptstyle{\parallel}}$ instead \cite{hubbard:pr1963}, where 
$\left(C_a^2 +  \tfrac{2}{9}C_d^2\right) = \tfrac{1}{3}\!\left(2C_{\perp}^2 + C_{\scriptscriptstyle{\parallel}}^2\right) $.

In order to compare with measurements, we also include contributions from the $^1$H-$^1$H dipole-dipole interactions, which separate into in{\it tra}molecular $T_{1R}$ and in{\it ter}molecular $T_{1T}$ relaxation. Details of the methodology behind the MD simulations of $T_{1R}$ and $T_{1T}$ can be found in Ref. \cite{singer:jmr2017}. The final expression for the total relaxation time is given by:
\begin{multline}
\frac{1}{T_{1}^{\alpha}} = \frac{1}{T_{1R}} + \frac{1}{T_{1T}}+ \frac{1}{T_{1J}^{\alpha}} = \\
\frac{10}{3} \Delta\omega_R^2  \tau_R +  \frac{10}{3} \Delta\omega_T^2 \tau_T +  2 \Delta\omega_{\alpha}^2 \tau_J,
\label{eq:T1sum}
\end{multline}
which takes on two different values $T_{1}^{km}$ and $T_{1}^{dm}$, depending on the model $T_{1J}^{km}$ and $T_{1J}^{dm}$ used. $\Delta\omega_{R,T}^2$ and $\tau_{R,T}$ are the second-moments and correlation times for in{\it tra}molecular ($R$) and in{\it ter}molecular ($T$), respectively. For comparison purposes, the square-root of the second-moments at 298 K and 349.1 bar (for instance) are 
$\Delta\omega_{R}/2\pi \simeq 23.7$ kHz, $\Delta\omega_{T}/2\pi \simeq 7.5$ kHz, and $\Delta\omega_{J}^{km}/2\pi \simeq 51.8$ kHz.
We also define the total relaxation from $^1$H-$^1$H dipole-dipole interactions $T_{1RT}$ as such:
\begin{align}
\frac{1}{T_{1RT}} = \frac{1}{T_{1R}} + \frac{1}{T_{1T}}
\label{eq:T1sumRT}
\end{align}
Given that the fast-motion regime applies, all of the above results for longitudinal relaxation $T_1$ apply equally to transverse relaxation $T_2$. As such, the subscript 2 has been dropped everywhere for clarity. 

\subsubsection{Distribution in correlation times} \label{sssc:DistributionInCorrelationTimes}

As shown in Fig. \ref{fg:GtDecay}(a), $G_{\!J}(t)$ deviates from single-exponential decay predicted by the Langevin model in Eq. \ref{eq:Langevin}. More specifically $G_{\!J}(t)$ has a more ``stretched" (i.e. multi-exponential) decay, which we quantify by inverting the following Laplace transform \cite{venkataramanan:ieee2002,song:jmr2002}: 
\begin{equation}
G_{\!J}(t) = \int\! P_J(\tau) \exp\bigl(-t/\tau\bigr) d\tau 
\label{eq:ILT}
\end{equation}
$P_J(\tau)$ (in units of s$^{-3}$) is the probability distribution function derived from the inversion. 
In the case of the Langevin sphere model, $P_J(\tau)$ is a delta-function at $\tau_J$, i.e. $P_J(\tau) = G_{\!J}(0) \, \delta(\tau - \tau_{J})$. However, as shown in Fig. \ref{fg:GtDecay}(a), $G_{\!J}(t)$ is always stretched (i.e. multi-exponential) to some degree, therefore $P_J(\tau)$ has a finite distribution. The decomposition of $G_{\!J}(t)$ into a sum of exponential decays is common practice  \cite{beckmann:prep1988}, where the more complex the molecule dynamics, the more terms are required. This justifies our general approach of decomposing $G_{\!J}(t)$ into a ``model free" sum of exponential decays in Eq. \ref{eq:ILT}, for the purposes of quantifying the departure from the Langevin sphere model.

The resulting $P_J(\tau)$ distributions, shown in Fig. \ref{fg:GtDecay}(b), were determined by using the discrete form of Eq. \ref{eq:ILT}, using 100 logarithmically-spaced $\tau$ bins ranging from 0.03 ps $\leq \tau \leq$ 30 ps, and a fixed regularization parameter of $10^{-2}$ \cite{venkataramanan:ieee2002,song:jmr2002}. Fig. \ref{fg:GtDecay}(b) shows that the mean correlation-times $\tau_J$ get longer with increasing $D_T$, and that the width of the distributions get somewhat larger with increasing $D_T$. In all cases, the widths are much larger than the delta function prediction from the Langevin model in Eq. \ref{eq:Langevin}.

The widths of the $P_J(\tau)$ distributions were then quantified using the following:
\begin{eqnarray}
\begin{aligned}
\mu_J &= \frac{1}{G_{\!J}(0) }\int\! P_J\!\left(\tau\right) \ln(\tau) \,d\tau, \\
\sigma_J^2 &= \frac{1}{G_{\!J}(0) }\int\! P_J\!\left(\tau\right) \left(\ln(\tau) - \mu_J \right)^2 d\tau.
\end{aligned} \label{eq:CvJ}
\end{eqnarray}
$\sigma_J$ is the standard deviation and $\mu_J$ is the mean of the variable $\ln(\tau)$. A natural logarithm in $\tau$ is used as the variable since the underlying $P_J(\tau)$ distributions are discrete and evenly spaced in $\ln(\tau)$. Also shown in Fig. \ref{fg:GtDecay}(b) is the Langevin model which predicts $\sigma_J = 0$ (i.e. a delta function), or $\sigma_J \simeq$ 0.038 due to regularization. Equivalent quantities were also derived for the in{\it tra}molecular $\sigma_R$ and in{\it ter}molecular $\sigma_T$ $^1$H-$^1$H dipole-dipole interactions.

\section{Results} \label{sc:Results}

\subsection{Simulation versus measurement} \label{ssc:SimulationVersusMeasurement}

\begin{figure}[!ht]
	\begin{center}
		\includegraphics[width=0.9\columnwidth,trim={0.2cm 0cm 0.8cm 0cm},clip]{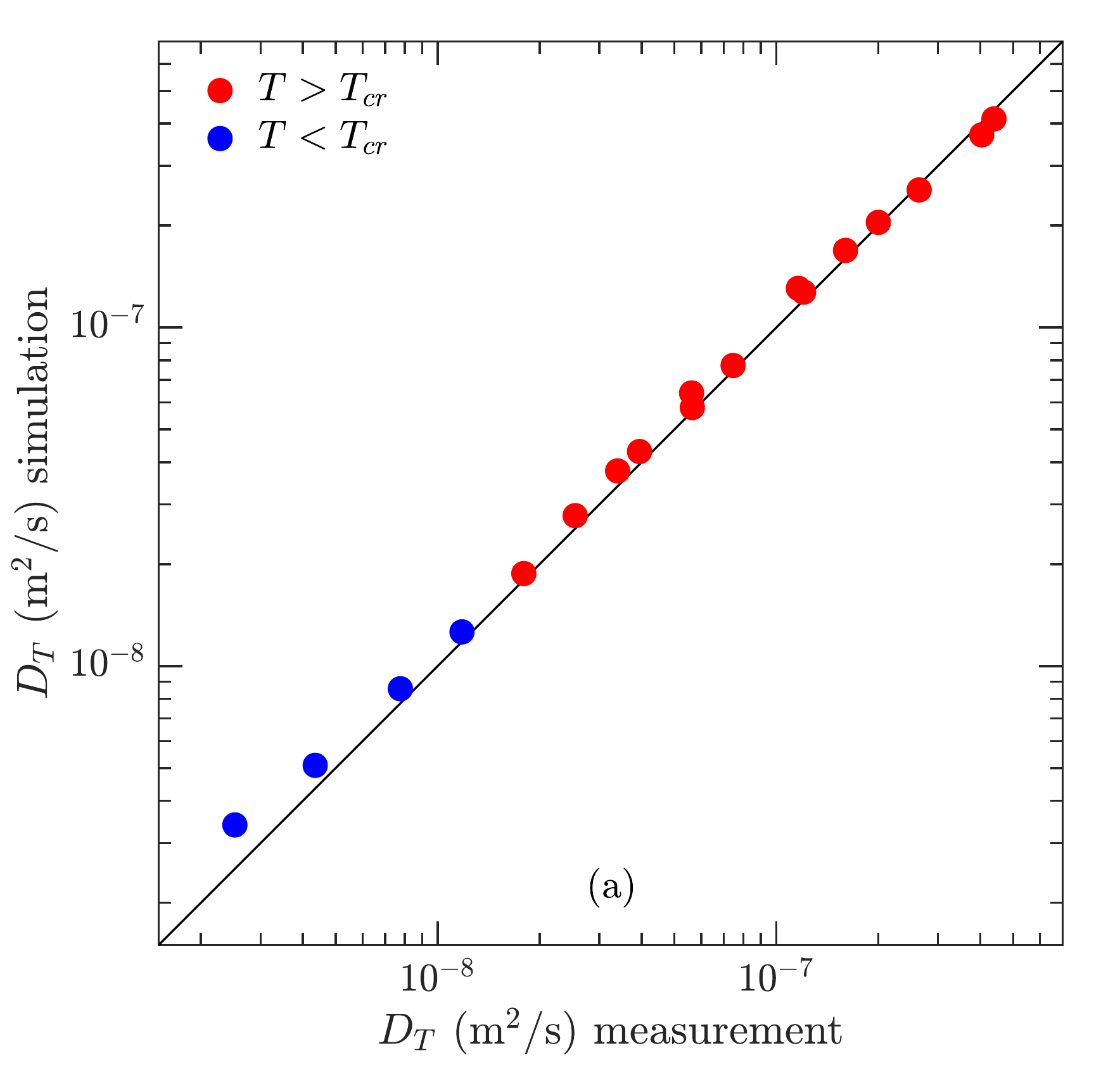}
		\includegraphics[width=0.9\columnwidth,trim={0.2cm 0cm 0.8cm 0cm},clip]{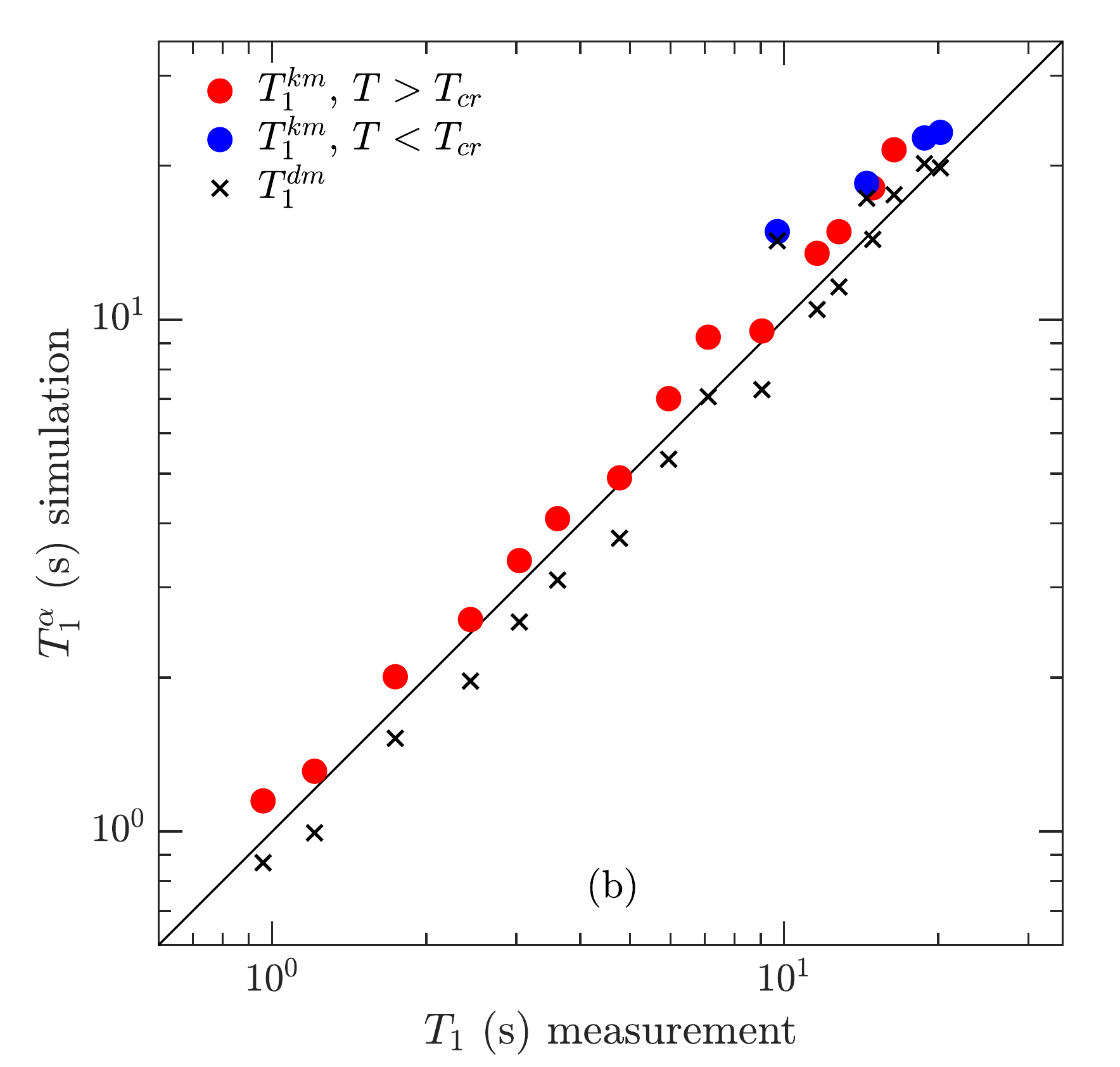}
	\end{center}
	\caption{(a) Cross-plot of simulated translational-diffusion coefficient $D_T$ on the $y$-axis, versus measured $D_T$ on the $x$-axis taken from Ref. \cite{gerritsma:ph1971a,gerritsma:ph1971b,oosting:ph1971a,oosting:ph1971b}, for both liquid state ($T< T_{cr}$), and supercritical or gas state ($T> T_{cr}$), listed in Table \ref{tb:1}. (b) Cross-plot of simulated total relaxation time $T_1^{\alpha}$ on the $y$-axis defined in Eq. \ref{eq:T1sum}, versus measured $T_1$ on the $x$-axis taken from Ref. \cite{gerritsma:ph1971a,gerritsma:ph1971b,oosting:ph1971a,oosting:ph1971b}, listed in Table \ref{tb:1}. The simulated $T_1^{\alpha}$ includes the kinetic model $T_{1}^{km}$ and the diffusion model $T_{1}^{dm}$.} 
	\label{fg:Correlations}
\end{figure}

The cross-plot of measured versus simulated translational-diffusion coefficient $D_T$ in Fig. \ref{fg:Correlations}(a) indicates a strong correlation coefficient $R^2  = 0.996$, and an average absolute deviation of $\delta_{abs} = 9.2$ \%, where $\delta_{abs}$ is defined as:
\begin{equation}
\delta_{abs} = \frac{1}{N}\sum\limits_{i=1}^{N}\left|\frac{Y_i - X_i}{X_i}\right|\times 100.
\label{eq:aad}
\end{equation}
$Y_i$ are the simulated values, $X_i$ are the measured quantities, and $N$ is the number of points. The deviation is noticeable at the lowest temperature in the liquid phase ($T<T_{cr}$), which may be due to the proximity of the liquid-vapor phase transition, and/or temperatures are low enough that nuclear quantum effects may be important (further investigations are beyond the scope of this work).


The cross-plot of measured $T_{1}$ versus simulated $T_{1}^{\alpha}$ (defined in Eq. \ref{eq:T1sum}) total-relaxation time in Fig. \ref{fg:Correlations}(b) indicates strong correlation coefficients of $R^2  = 0.982$ for $T_{1}^{km}$, and $R^2  = 0.954$ for $T_{1}^{dm}$, while $\delta_{abs}$ varies as a function of measured $T_{1}$. At the higher $D_T$ (i.e. lower $\rho$) end where the measured $T_{1} < 5 $ s, the deviation is lower for the kinetic model $T_{1}^{km}$ ($\delta_{abs}$ = 11.0 \%) than for the diffusion model $T_{1}^{dm}$ ($\delta_{abs}$ = 15.8 \%). This is expected given that the kinetic model is more appropriate for fluids at high $D_T$ and low $\rho$. 

At the lower $D_T$ (i.e. higher $\rho$) end where the measured $T_{1} > 5 $ s {\it and} $T > T_{cr}$ (i.e. still in the supercritical phase), the deviation is lower for the diffusion model $T_{1}^{dm}$ ($\delta_{abs}$ = 8.6 \%) than for the kinetic model $T_{1}^{km}$ ($\delta_{abs}$ = 19.7 \%). This is expected given that the diffusion model is more appropriate for fluids at low $D_T$ and high $\rho$. 

In the liquid phase where $T < T_{cr}$, $^1$H-$^1$H dipole-dipole relaxation begins to dominate over the spin-rotation interaction, and the two interpretations $T_{1}^{km}$ and $T_{1}^{dm}$ become comparable. Nevertheless, the deviation is still lower for the diffusion model $T_{1}^{dm}$ ($\delta_{abs}$ = 18.9 \%) than for the kinetic model $T_{1}^{km}$ ($\delta_{abs}$ = 29.1 \%). The source of the deviation is from the dipole-dipole contribution $T_{1RT}$, more specifically from the dominating in{\it ter}molecular contribution $T_{1T}$. This systematic deviation is potentially due to the proximity of the liquid-vapor phase transition, and/or temperatures are low enough that nuclear quantum effects may be important (further investigations are beyond the scope of this work).

Another potential systematic error are the experimental uncertainties in the coupling constants $C_a$ and $C_d$ \cite{yi:jcp1971} listed in Section \ref{sc:Methodology}. The maximum and minimum deviations reported in \cite{yi:jcp1971} result in a $\pm$2.7 \% uncertainty in $T_{1J}^{km}$, and a $\pm$3.3 \% uncertainty in $T_{1J}^{dm}$. These uncertainties are not insignificant in the above analysis, and should be taken into consideration when interpreting the simulations. There may also be uncertainty in the moment of inertia $I$ for methane \cite{herzberg:book}.

Yet another potential systematic error in the measurements is the presence of dissolved oxygen \cite{shikhov:amr2016}. O$_2$ is paramagnetic, which would tend to shorten the measured $T_1$ compared to simulations. Given the large values of measured $T_1 \leq$ 20 s, any trace amounts of oxygen could affect the results.

\subsection{Correlation times} \label{ssc:CorrelationTimes}

\begin{figure}[!ht]
	\begin{center}
		\includegraphics[width=0.9\columnwidth,trim={0.2cm 1.8cm 0.4cm 0cm},clip]{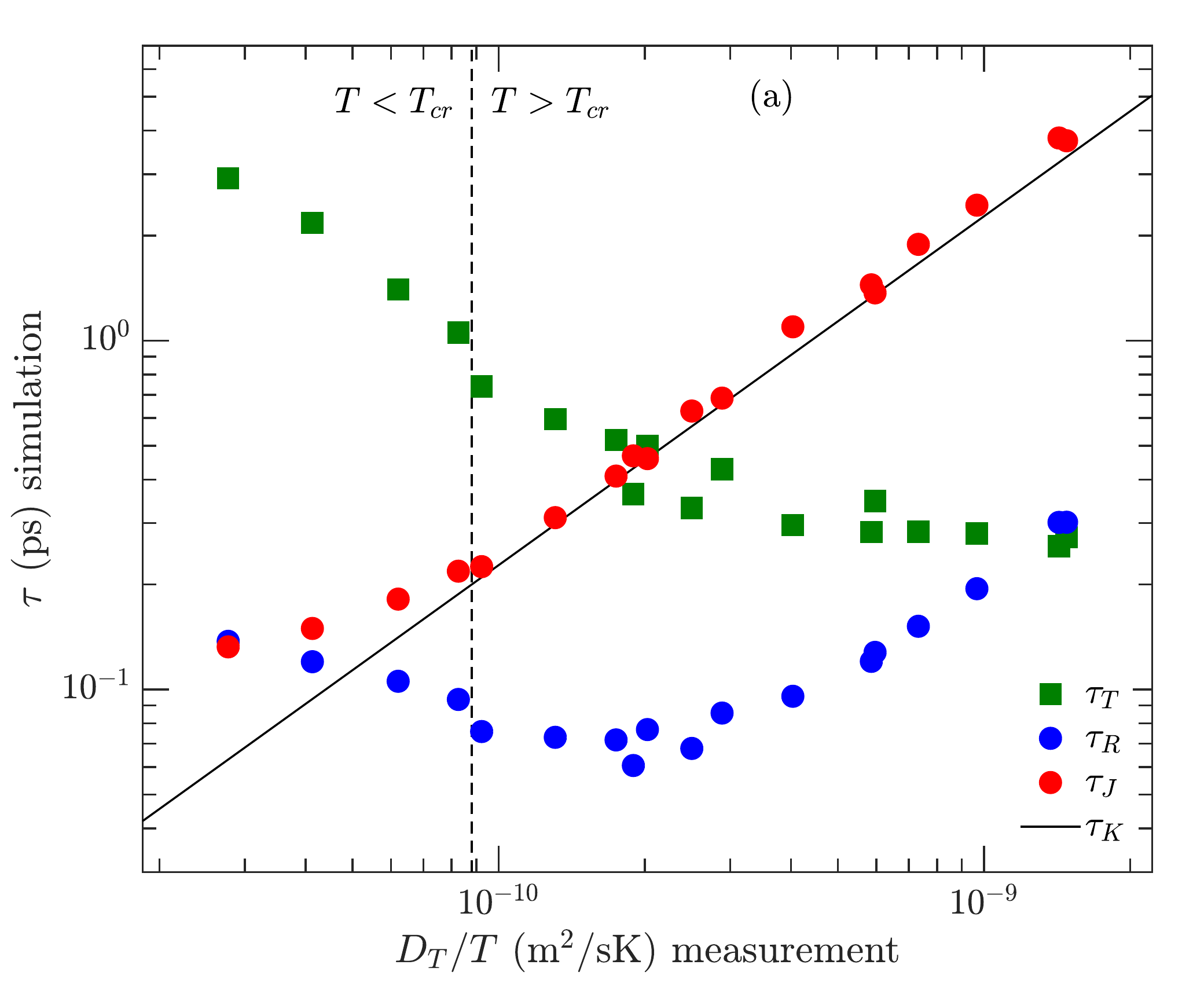}  
		\includegraphics[width=0.9\columnwidth,trim={0.2cm 0cm 0.4cm 0.4cm},clip]{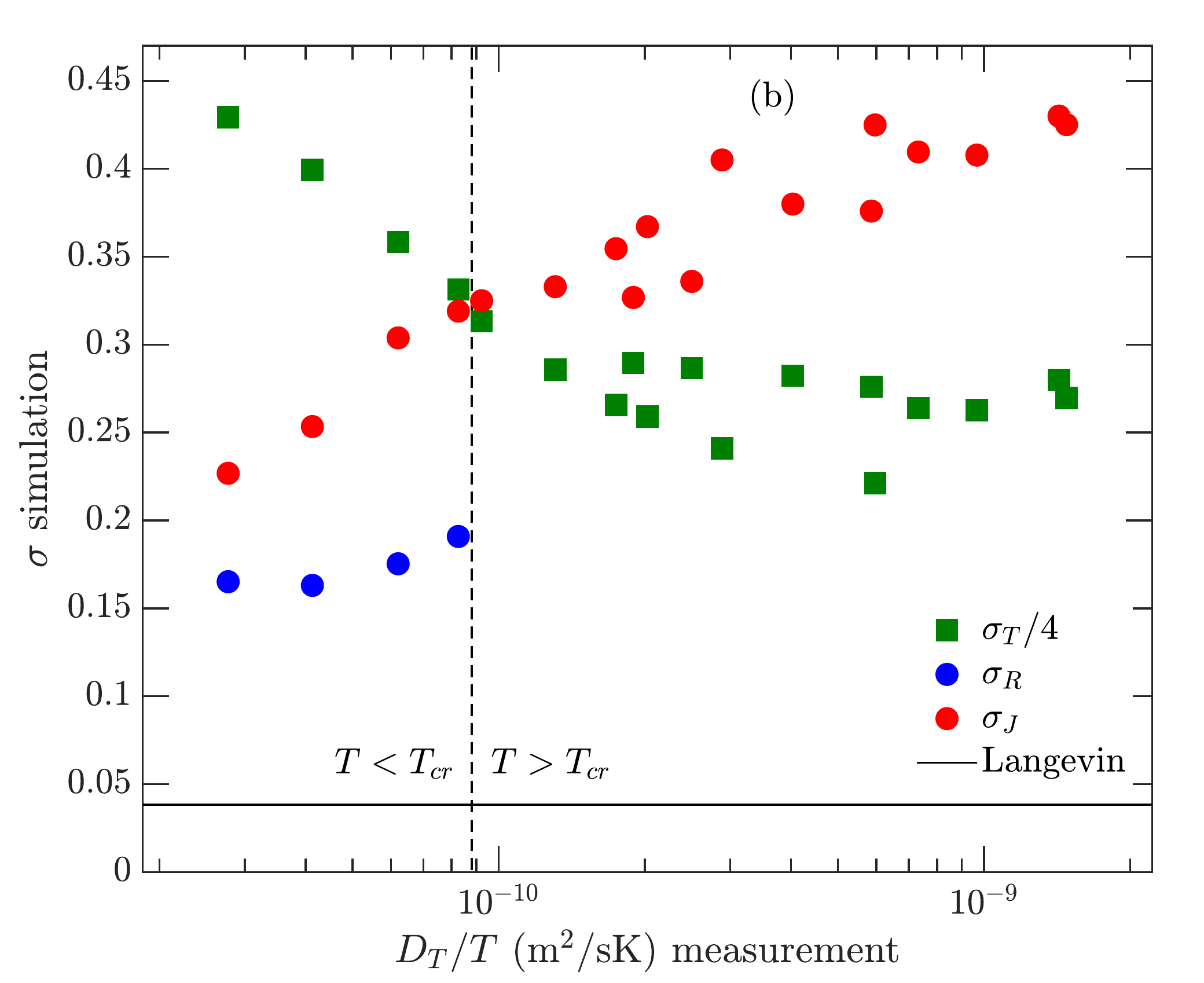}
	\end{center}
	\caption{(a) MD simulated correlation times for in{\it ter}molecular $\tau_T$ and in{\it tra}molecular $\tau_R$ $^1$H-$^1$H dipole-dipole interactions, spin-rotation interaction $\tau_J$ (Eq. \ref{eq:TauJ}), and mean collision time $\tau_K$ (Eq. \ref{eq:Lambda}), plotted against measured $D_T/T$. (b) Simulation results for the standard deviation $\sigma_J$ (Eq. \ref{eq:CvJ}) in correlation times, determined from the $P_J(\tau)$ distributions (Eq. \ref{eq:ILT}) in Fig. \ref{fg:GtDecay}(b), plotted against measured $D_T/T$. Also shown is the Langevin model which predicts a delta function $\sigma_J = 0$, or $\sigma_J \simeq$ 0.038 due to regularization. Equivalent quantities $\sigma_R$ and $\sigma_T$ for $^1$H-$^1$H dipole-dipole interactions are also shown, where $\sigma_T/4$ is plotted for clarity.} 
	\label{fg:Taus}
\end{figure}

A summary of the simulated correlation times is shown in Fig. \ref{fg:Taus}(a). The spin-rotation correlation time $\tau_J$ shows a monotonic increase with increasing $D_T/T$ (translational-diffusion divided by absolute temperature). At high $D_T/T$, i.e. in the supercritical/gas phase ($T > T_{cr}$), $\tau_J$ is consistent with the kinetic collision time $\tau_K$ defined as:
\begin{align}
D_T &= \frac{1}{3} \lambda \bar{v}, \quad
\bar{v} = \sqrt{\frac{8kT}{\pi M}},\label{eq:Lambda}  \\
\tau_K &= \frac{\lambda}{\bar{v}} = \frac{3D_T}{\bar{v}^2} =  \frac{3\pi M}{8k} \frac{ D_T}{T},
\nonumber
\end{align}
where $M = 2.664 \times 10^{-26}$ kg is the mass of the methane molecule, and $\bar{v}$ is the mean thermal velocity. More specifically, the correlation between $\tau_J $ and $ \tau_K$ is strong $R^2 = 0.997$ and the absolute deviation is $\delta_{abs} = 7.7$ \% in this region. Given the compelling observation $\tau_J \simeq \tau_K$ and the relation $\tau_K \propto D_T/T$ in Eq. \ref{eq:Lambda} motivates using $D_T/T$ for the $x$-axis in Fig. \ref{fg:Taus}. The relation $\tau_J \simeq \tau_K$ is indicative of the ``strong collision" regime \cite{bloom:cjp1967b,oosting:ph1971a}, and is used below to infer a new relation for $T_{1J}^{km}$. As expected, the relation $\tau_J \simeq \tau_K$ breaks down in the liquid phase ($T < T_{cr}$) where the diffusion model is more appropriate.

In the case of spherical molecules, the following simple theoretical predictions exist for relations between the in{\it tra}molecular correlation time $\tau_R$ and spin-rotation correlation time $\tau_J$:
\begin{align}
\tau_R &= \frac{\tau_{J}}{2l\!+\!1} = \frac{\tau_{J}}{5} \,\,\,\,[{\rm k.m.}], \label{eq:Bloom} \\
\tau_R \tau_J&= \frac{I}{l(l\!+\!1)kT} = \frac{I}{6kT} \,\,\,\,[{\rm d.m.}]. \label{eq:Hubbard}
\end{align}
In the kinetic model ($T > T_{cr}$) for the supercritical/gas phase, Eq. \ref{eq:Bloom} states that $\tau_R = \tau_J/5$ \cite{bloom:cjp1967b,mcclung:JCP1969}, where $l=2$ is the rank of rotational diffusion tensor for hard-spheres. As shown in Fig. \ref{fg:Taus}(a) for $T > T_{cr}$, the relation $\tau_R \propto \tau_J$ is indeed found, however Eq. \ref{eq:Bloom} overestimates $\tau_R$ by a factor $\simeq 2.5$. In the diffusion model ($T < T_{cr}$) for liquids, Eq. \ref{eq:Hubbard} \cite{hubbard:pr1963} makes use of the Stokes-Einstein relation for hard-spheres, which predicts that rotational-diffusion $D_R = 1/l(l\!+\!1)\tau_R = 1/6\tau_R$ and translational-diffusion $D_T$ are related by $D_T = 4a^2D_R/3$ \cite{hubbard:pr1963}, where $a$ is the methane radius. At the lowest $D_T$ value the relation $\tau_R \propto 1/\tau_J$ is indeed found, however Eq. \ref{eq:Hubbard} underestimates $\tau_R$ by a factor $\simeq 2.5$. 

In the liquid phase ($T < T_{cr}$), the in{\it ter}molecular correlation time $\tau_T$ decreases with increasing $D_T$, in general accordance with the Stokes-Einstein relation for hard-spheres $\tau_D = \tfrac{5}{2} \tau_T = 2a^2/D_T$ \cite{singer:jmr2017}. A transport radius of $a \simeq 1.0 \, {\rm \AA}$ can be inferred at the lowest $D_T$ value, which is consistent with the C-H internuclear distance of  $a \simeq 1.093 \, {\rm \AA}$ deduced from the relation $I = \tfrac{8}{3} m_H a^2$ \cite{herzberg:book}, where $m_H$ is the $^1$H mass. At high $D_T$, i.e. in the supercritical/gas phase ($T > T_{cr}$), the Stokes-Einstein relation breaks down, and $\tau_T$ becomes independent of $D_T$.

A summary of the simulated standard deviations $\sigma$ in correlation-times is shown in Fig. \ref{fg:Taus}(b). For spin-rotation, $\sigma_J$ increases by a factor of two from the lowest $D_T$ to the highest $D_T$, indicating a larger distribution in correlation times for the gas phase. In the case of  in{\it tra}molecular dipole-dipole, $\sigma_R$ shows a low value in the liquid phase, and a possible increase with increasing $D_T$. In the supercritical/gas phase, $G_R(t)$ shows signs of oscillations at early times $t \lesssim 0.3$ ps, and then decays monotonically for $t \gtrsim 0.3$ ps. As such, $\sigma_R$ is not computed for $T>T_{cr}$. Note that oscillations in $G_R(t)$ for $T>T_{cr}$ were previously predicted using the extended diffusion model \cite{mcclung:JCP1971,mcclung:AMRIP1977}.

Meanwhile, $\sigma_T$ decreases with increasing $D_T$, which is the opposite trend to $\sigma_J$ and $\sigma_R$. At high $D_T$, $\sigma_T$ plateaus to the $\sigma_T \simeq 1.25$ (note that Fig. \ref{fg:Taus}(b) displays $\sigma_T/4$ for clarity), which is consistent with the inherent multi-exponential value of the Torrey hard-sphere model \cite{torrey:pr1953}. 

\begin{figure}[!ht]
	\begin{center}
		\includegraphics[width=0.9\columnwidth,trim={0.2cm 1.8cm 0.4cm 0cm},clip]{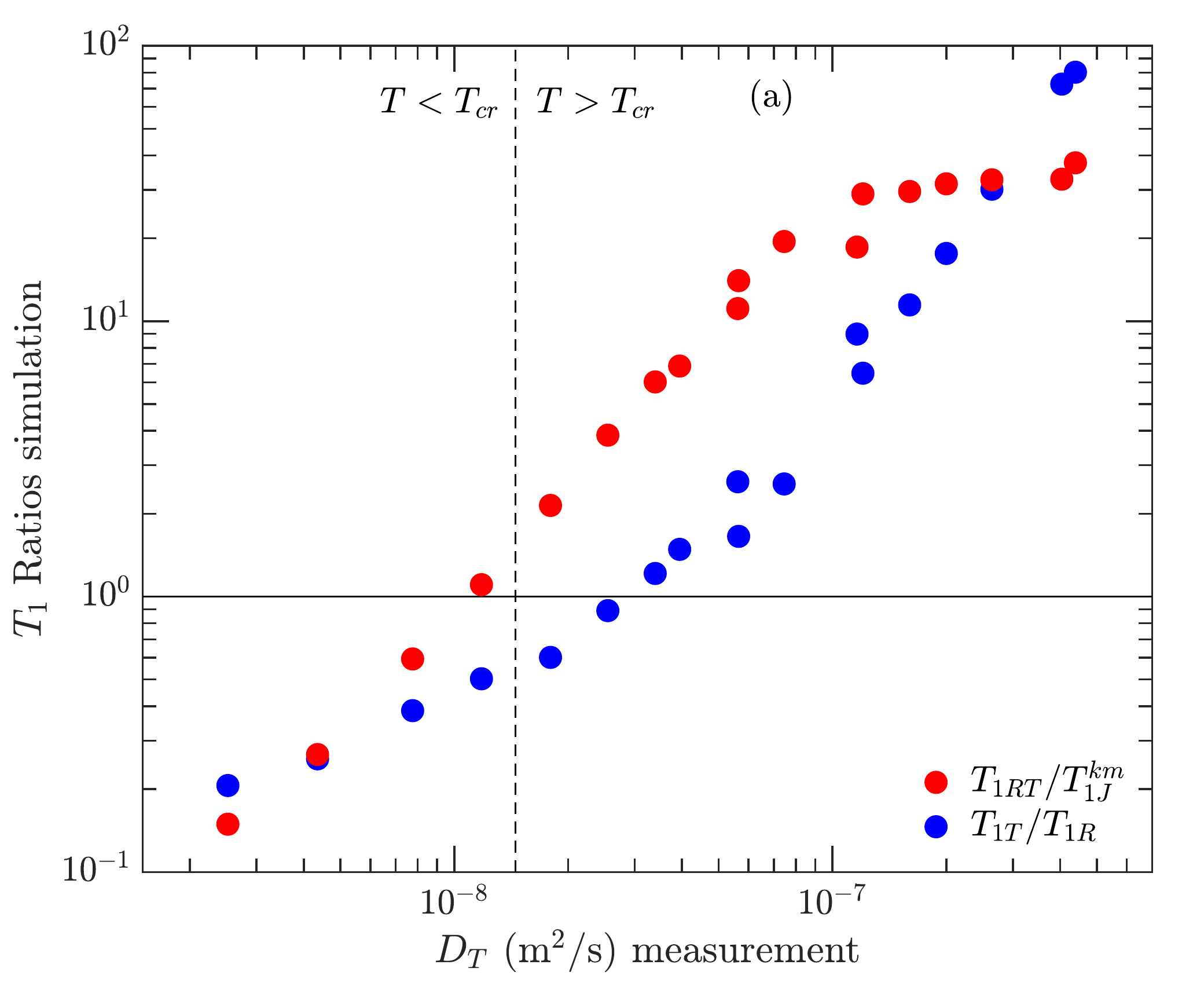}  
		\includegraphics[width=0.9\columnwidth,trim={0.2cm 0cm 0.4cm 0.4cm},clip]{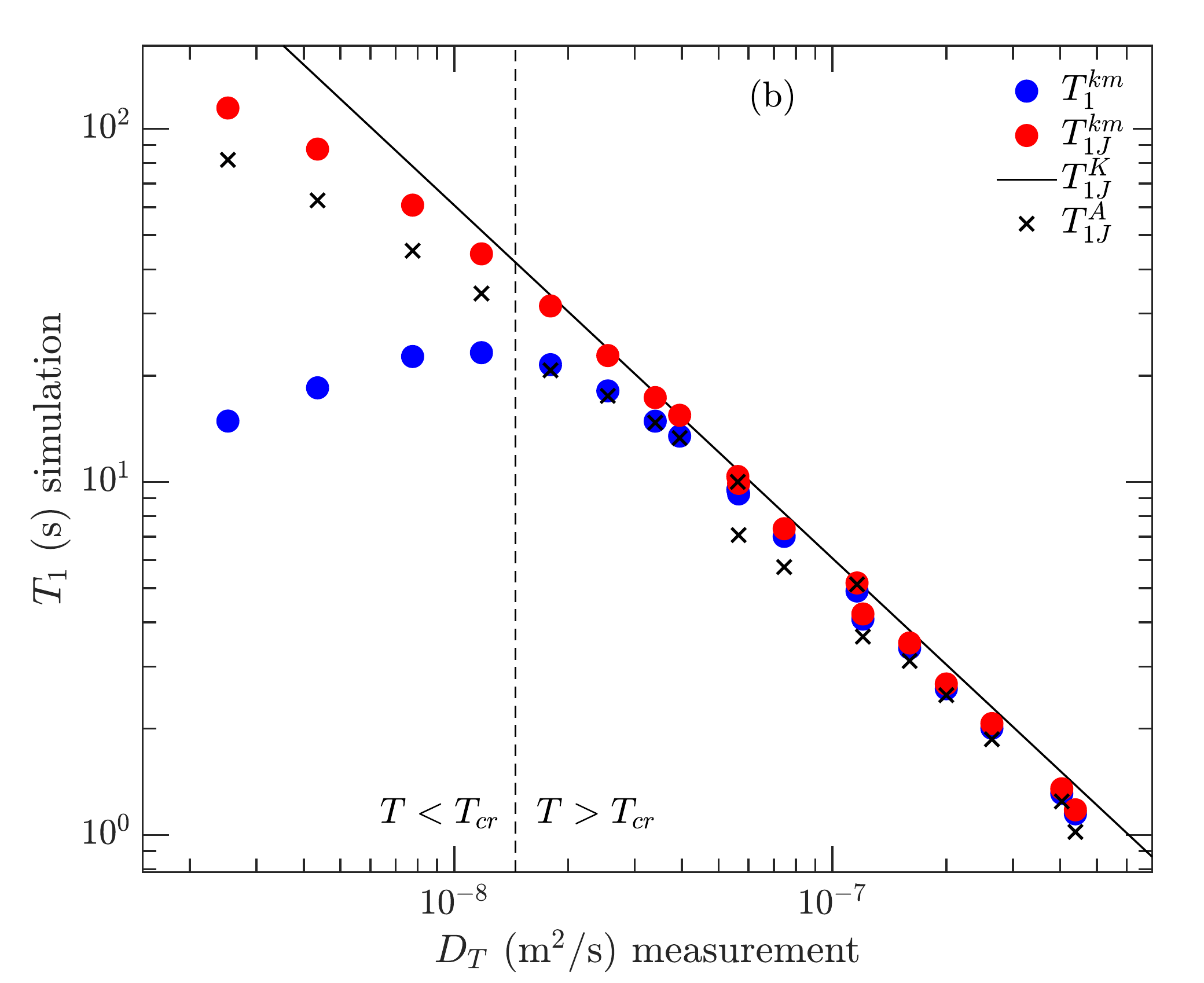}  
	\end{center}
	\caption{(a) Ratio of simulated $T_1$ relaxation times to display relative strengths of the interactions; including, ratio of total dipole-dipole $T_{1RT}$ (Eq. \ref{eq:T1sumRT}) to  spin-rotation $T_{1J}^{km}$ (Eq. \ref{eq:T1Jkm}), and ratio of in{\it ter}molecular dipole-dipole $T_{1T}$ to in{\it tra}molecular dipole-dipole $T_{1R}$, plotted against measured $D_T$. (b) Simulated spin-rotation relaxation time $T_{1J}^{km}$ (Eq. \ref{eq:T1Jkm}), empirical expression $T_{1J}^{A}$ (Eq. \ref{eq:T1lo}), new prediction $T_{1J}^{K}$ (Eq. \ref{eq:T1lambda}), and total relaxation time $T_1^{km}$ (Eq. \ref{eq:T1sum}), plotted against measured $D_T$.} 
	\label{fg:RatiosLambda}
\end{figure}

As stated in Eqs. \ref{eq:T1Jkm} and \ref{eq:T1Jdm}, the longer the spin-rotation correlation-time $\tau_J$, the shorter the relaxation time $T_{1J}$, i.e. the more significant the relaxation mechanism. The same is true for the $^1$H-$^1$H dipole-dipole mechanism \cite{singer:jmr2017}. It is therefore informative to compare the ratio of relaxation times between different mechanisms in order to determine which mechanism is dominant. Fig. \ref{fg:RatiosLambda}(a) shows the ratio $T_{1RT}/T_{1J}^{km}$, where $T_{1RT}$ is the total dipole-dipole relaxation defined in Eq. \ref{eq:T1sumRT}. A large value $T_{1RT}/T_{1J}^{km} \gg 1$ indicates that spin-rotation dominates over dipole-dipole, as found at high $D_T$ in the supercritical/gas phase ($T > T_{cr}$). Meanwhile a small value $T_{1RT}/T_{1J}^{km} \ll 1$ indicates that dipole-dipole dominates over spin-rotation, as found at low $D_T$ in the liquid phase ($T < T_{cr}$). 

A similar analysis can be made between the in{\it tra}molecular and the in{\it ter}molecular dipole-dipole interactions. A large value $T_{1T}/T_{1R} \gg 1$ indicates that in{\it tra}molecular dominates over in{\it ter}molecular, as found at high $D_T$ in the supercritical/gas phase ($T > T_{cr}$). Meanwhile a small value $T_{1T}/T_{1R} \ll 1$ indicates that in{\it ter}molecular dominates over in{\it tra}molecular, as found at low $D_T$ in the liquid phase ($T < T_{cr}$). The most likely reason for this is that $\tau_T \gg \tau_R$ in the liquid phase ($T < T_{cr}$). Meanwhile in the supercritical/gas phase ($T > T_{cr}$), even though $\tau_T \simeq \tau_R$, the in{\it ter}molecular second-moment $\Delta\omega_T^2 \propto \rho$ \cite{singer:jmr2017}, and therefore $\Delta\omega_T^2$ decreases with increasing $D_T$ (i.e. decreasing $\rho$).

\subsection{New Kinetic model} \label{ssc:KineticModel}

Building on the observation that $\tau_J \simeq \tau_K$ in the supercritical/gas phase ($T > T_{cr}$) (see Fig. \ref{fg:Taus}(a)), and the observation that spin-rotation dominates over dipole-dipole $T_{1RT}/T_{1J}^{km} \gg 1$ in this region (Fig. \ref{fg:RatiosLambda}(b)), it is informative to infer a new relation for $T_{1J}^{km} \simeq T_{1}^{km}$ in the supercritical/gas phase. Assuming  $\tau_J = \tau_K$ in Eq. \ref{eq:T1Jkm}, and using the definition for $\tau_K$ in Eq. \ref{eq:Lambda} results in the following prediction:
\begin{align}
\frac{1}{T_{1J}^{K}} = \frac{3\pi I M}{4 \hbar^2} \left(C_a^2 +  \tfrac{4}{45}C_d^2\right) D_T \,\,\,\,[{\rm k.m.}].
\label{eq:T1lambda}
\end{align}
The new prediction in the kinetic regime $T_{1J}^K$ is plotted in Fig. \ref{fg:RatiosLambda}(b) against the simulated $T_{1J}^{km}$ and the measured $D_T$. In the supercritical/gas phase ($T > T_{cr}$), the correlation coefficient between $T_{1J}^{km}$ and $T_{1J}^K$ is found to be $R^2 = 0.998$, and the absolute deviation is found to be $\delta_{abs} = 7.7$ \%, without any adjustable parameters in the derivation of $T_{1J}^K$ in Eq. \ref{eq:T1lambda}. Moreover, there are no assumptions about an effective molecular radius. In the liquid phase ($T < T_{cr}$), $T_{1J}^{km}$ starts to deviate from the kinetic model $T_{1J}^K$, as expected. Likewise, the total relaxation $T_{1}^{km}$ deviates from $T_{1J}^K$ due to the $T_{1RT}$ contribution, as expected. 

Also shown in Fig. \ref{fg:RatiosLambda}(b) is the empirical relation $T_{1J}^A$ from Eq. \ref{eq:T1lo}, where $A = 6.37 \times 10^{-6}$ \cite{lo:SPE2002} in units of temperature $T$ (K), density $\rho$ (g/cm$^3$), and $T_{1J}^A$ (s). In the supercritical/gas phase ($T > T_{cr}$), the correlation between $T_{1J}^A$ and $T_{1J}^{K}$ is strong $R^2 = 0.957$, while the absolute deviation is $\delta_{abs} \simeq 21.2 $ \%, for possibly the same reasons as discussed in Section \ref{ssc:SimulationVersusMeasurement}. Eq. \ref{eq:T1lo} and Eq. \ref{eq:Langevin} imply that the $T$ and $\rho$ dependence of $T_{1J}$ comes entirely from $D_T \propto T^{3/2}\!/\!\rho$. This is confirmed in the supercritical/gas phase ($T > T_{cr}$), where the correlation coefficient between measured $D_T$ and measured $T^{3/2}\!/\!\rho$ in Table \ref{tb:1} is strong $R^2 = 0.991$.

\section{Conclusions}\label{sc:Conclusions}

We develop a minimization technique to compute the angular-velocity for non-rigid spherical molecules, which is used to simulate the autocorrelation function $G_{\!J}(t)$ for the spin-rotation interaction of methane, over a wide range of densities $\rho$ and temperatures $T$, spanning the liquid ($T<T_{cr}$), and supercritical/gas ($T>T_{cr}$) phases. The Langevin model predicts that $G_{\!J}(t)$ should decay with a single-exponential function with correlation time $\tau_J$. However, inverse Laplace transforms of $G_{\!J}(t)$ indicate a distribution in correlation times $\tau$, with a standard deviation $\sigma_J$ (i.e. width) which increases with increasing $D_T$ (i.e. decreasing $\rho$).

$T_{1J}^{\alpha}$ is derived from $G_{\!J}(t)$ using the kinetic model ``{\it km}" for gases ($T_{1J}^{km}$) \cite{bloom:cjp1967b}, and the diffusion model ``{\it dm}" for liquids ($T_{1J}^{dm}$)  \cite{hubbard:pr1963}. The total relaxation time $T_{1}^{km}$ shows better agreement with measurements for $T_{1} < $ 5 s, with an absolute deviation of $\delta_{abs} = 11.0$ \% in this region. $T_{1}^{dm}$ shows better agreement with measurements for $T_{1} > $ 5 s (and $T>T_{cr}$), with an absolute deviation of $\delta_{abs} = 8.6$ \% in this region. Uncertainties in the measured spin-rotation coupling-constants may contribute to these deviations. Meanwhile the simulated $D_T$ agree well with measurements, without any adjustable parameters in the interpretation of the simulations.

MD simulations of the in{\it tra}molecular and in{\it ter}molecular $^1$H-$^1$H dipole-dipole relaxation are computed at the same state-points. $T_{1J}^{km}$ is shown to dominate over the total dipole-dipole relaxation $T_{1RT}$ at high $D_T$, while the opposite is found at low $D_T$. 

The predicted relations between the in{\it tra}molecular correlation-time $\tau_R$ and the spin-rotation correlation-time $\tau_{J}$ is tested, both in the liquid ($T<T_{cr}$) and the supercritical/gas phase ($T<T_{cr}$). At the highest $D_T$ in the supercritical/gas phase, the relation $\tau_R = \tau_{J}/5$ \cite{bloom:cjp1967b,mcclung:JCP1969} is found to hold within a factor $\simeq 2.5$. At the lowest $D_T$ in the liquid phase, the relation $\tau_R \tau_J = I/6kT$ \cite{hubbard:pr1963} is also found to hold within a factor $\simeq 2.5$.

In the supercritical/gas phase ($T>T_{cr}$), $\tau_{J}$ is found to agree with the kinetic collision time $\tau_{K}$, with an absolute deviation of $\delta_{abs} = 7.7$ \% in this region. Given this compelling finding, a new expression for the spin-rotation relaxation  $1/T_{1J}^{K} \propto D_T $ is inferred without any adjustable parameters, and shows a strong correlation $R^2 = 0.957$ with the previously reported empirical finding $1/T_{1J}^{A} \propto T^{3/2}\!/\rho$.

\section*{Acknowledgments} 

This work was funded by the Rice University Consortium on Processes in Porous Media, and the American Chemical Society Petroleum Research Fund [ACS-PRF-58859-ND6]. We gratefully acknowledge the National Energy Research Scientific Computing Center, which is supported by the Office of Science of the U.S. Department of Energy [DE-AC02-05CH11231], for HPC time and support. We also gratefully acknowledge the Texas Advanced Computing Center (TACC) at The University of Texas at Austin (URL: http://www.tacc.utexas.edu) for providing HPC resources, and Zeliang Chen for his assistance.



\end{document}